\documentclass[ amsmath,amssymb,
reprint,%
]{revtex4-1}

\usepackage{graphicx}
\usepackage{dcolumn}
\usepackage{bm}

\usepackage[utf8]{inputenc}
\usepackage[T1]{fontenc}
\usepackage{mathptmx}
\usepackage{etoolbox}
\usepackage{changes}
\usepackage{verbatim}
\usepackage{color}

\usepackage[normalem]{ulem}
\usepackage{booktabs}
\usepackage{ulem}
\usepackage{soul}
\definecolor{reviewer1}{RGB}{0,0,0} 
\definecolor{reviewer2}{RGB}{0,0,0} 
\definecolor{reviewerboth}{RGB}{0,0,0} 
\definecolor{reviewint}{RGB}{0,0,0} 
\definecolor{SS}{RGB}{0,0,0} 

\makeatletter
\def\@email#1#2{%
	\endgroup
	\patchcmd{\titleblock@produce}
	{\frontmatter@RRAPformat}
	{\frontmatter@RRAPformat{\produce@RRAP{*#1\href{mailto:#2}{#2}}}\frontmatter@RRAPformat}
	{}{}
}
\makeatother

\begin{document}
	
	\preprint{xxx}
	
	\title[Magnetic Force Microscopy: High Quality-Factor Two-Pass Mode]{Magnetic Force Microscopy: High Quality-Factor Two-Pass Mode}
	
	\author{Christopher Habenschaden}
	\homepage{https://www.ptb.de/cms/en/ptb/fachabteilungen/abt2/fb-25/ag-252.html}
	
	\author{Sibylle Sievers}
	\affiliation{ 
		Physikalisch-Technische Bundesanstalt (PTB), 38116 Braunschweig, Germany
	}
	
	\author{Alexander Klasen}
	\author{Andrea Cerreta}
	\affiliation{
		Park Systems Europe GmbH, 68199 Mannheim, Germany
	}
	
	\author{Hans Werner Schumacher}%
	\affiliation{ 
		Physikalisch-Technische Bundesanstalt (PTB), 38116 Braunschweig, Germany 
	}

\date{October 24, 2024}

\begin{abstract}

    Magnetic force microscopy (MFM) is a well-established technique in scanning probe microscopy that allows for the imaging of magnetic samples with a spatial resolution of tens of nm and stray fields down to the mT range. The spatial resolution and field sensitivity can be significantly improved by measuring in vacuum conditions. This improvement originates from the higher quality-factor (Q-factor) of the cantilever's oscillation in vacuum compared to ambient conditions. However, while high Q-factors are desirable as they directly enhance the magnetic measurement signal, they pose a challenge when performing standard MFM two-pass (lift) mode measurements. At high Q-factors, amplitude-based topography measurements become impossible, and the MFM phase response behaves non-linearly. Here, we present a modified two-pass mode implementation in a vacuum atomic force microscope (AFM) that addresses these issues. By controlling the Q-factor in the first pass and using a phase-locked loop (PLL) technique in the second pass, high Q-factor measurements in vacuum are enabled. Measuring the cantilever's frequency shift instead of the phase shift eliminates the issue of emerging nonlinearities. The improvements in MFM \textcolor{SS}{signal-to-noise ratio} are demonstrated using a nano-patterned magnetic sample. The elimination of non-linear response is highlighted through measurements performed on a well-characterized multilayer reference sample. \textcolor{SS}{Finally, we discuss a technique that avoids topography-induced artifacts by following the average sample slope. The newly developed, sensitive, and distortion-free high quality-factor two-pass mode has the potential to be widely implemented in commercial setups, facilitating high-resolution MFM measurements and advancing studies of modern magnetic materials.}

    \noindent --------- 
    
    \noindent The following article has been accepted by Review of Scientific Instruments. After it is published, it will be found at \url{https://pubs.aip.org/aip/rsi}. Copyright 2024 Author(s). This article is distributed under a Creative Commons Attribution-NonCommercial-NoDerivs 4.0 International (CC BY-NC-ND) License.

\end{abstract}

\maketitle

\section{\label{sec:intro}Introduction}
    
    Magnetic force microscopy (MFM) is a widely accessible, user-friendly, and powerful tool for characterizing materials with magnetic micro- and nanostructures. \textcolor{reviewer1}{Most commonly, MFM setups today} detect the interaction between a magnetically coated tip on an oscillating cantilever and the sample, mapping the emanating stray fields. \textcolor{reviewer1}{This is known as dynamic mode; however, static modes are also available.}
    
    Initially, the development of MFM was driven by the industry's need to analyze and characterize magnetic data storage media \cite{Babcock1994}. However, modern magnetic materials, which are the focus of current research, reveal new challenges for characterization. Magnetic data storage is evolving not only by pushing the density of magnetic data to its physical limits, but also by exploring novel data storage methods \cite{Morris2021}. Research topics now include studies on multilayers for spintronic applications, vortex structures like skyrmions \cite{Luo2021}, and 2D materials, increasingly focusing on very low stray fields and nanoscale structures \cite{Liu2023}. As a result, MFM techniques must evolve \textcolor{SS}{toward higher sensitivity and higher spatial resolution} to meet these new demands.

    \textcolor{reviewer1}{One way to achieve higher resolution is through careful selection of the tip. Various tip geometries, with different coatings tailored for specific operation modes, are available. Recent developments have shifted from coated tips to ultra-sharp fabricated spike-tips, nanowires, and filled carbon nanotubes, enabling high-resolution measurements \cite{Choi2010, Belova2012, Wolny2011, Vock2010}. In this work, a different approach is discussed:}

    The spatial resolution and field sensitivity of MFM can be significantly enhanced by performing measurements under vacuum conditions \cite{Feng2022}. This improvement is due to the higher cantilever quality factor (Q-factor) in vacuum, which directly enhances force sensitivity and, consequently, the signal-to-noise ratio (SNR) of the measurement. However, the commonly used two-pass mode for MFM measurements faces certain limitations under vacuum conditions. In two-pass mode, each line of the sample is scanned twice. In the first pass, the topography of the sample is detected by controlling the amplitude of the cantilever’s oscillation in a feedback loop that varies the tip-sample distance. In the second pass, the cantilever is lifted, and the phase change of its oscillation is measured, which is linked to the sample’s magnetic stray fields.
    
    \textcolor{SS}{In vacuum, due to the high Q-factors, only a small amount of energy is dissipated per oscillation cycle, requiring minimal driving force. While this significantly enhances sensitivity to external forces, it also makes controlling the oscillation more challenging. During the first pass, where the tip is brought close to the surface to map the topography, this means that the external forces can overpower the total restoring force of the oscillation, potentially causing the tip to crash. \textcolor{reviewer1}{Specifically, the total restoring force at the lower turning point of the oscillation must exceed the attractive forces between the tip and the sample \cite{Meyer2021}.} High Q-factors can thus result in tip crashes, preventing stable operation.}
    
    \textcolor{SS}{During the second pass, where the phase change is measured, high Q-factors lead to long integration times, as bandwidth and quality factor are interdependent. This can be mitigated by artificially lowering the Q-factor (Q-control); however, this compromises sensitivity. Additionally, as discussed below, even moderately increased Q-factors can cause non-linearities, such as the loss of linearity between phase shift and the underlying magnetic field, which may introduce severe distortions in the MFM image. This makes high Q-factor phase shift data prone to misinterpretation. Both issues -- long measurement times and signal non-linearities -- can be addressed by measuring frequency shift data instead of phase shift data during the lifted pass.}
    
    \textcolor{SS}{One approach discussed in the literature to implement frequency shift measurements is to avoid the topography pass altogether.} \textcolor{reviewer1}{This can be achieved by carefully determining the sample's slope in the x- and y-directions in non-contact mode beforehand and performing solely a lifted scan, as implemented in NanoScan high-resolution MFM setups, see, for example, Refs. [\onlinecite{Wolny2010, Samad2021, Arekapudi2021}].} Another approach is bimodal magnetic force microscopy with capacitive tip-sample distance control, as described in Refs. [\onlinecite{Schwenk2015, Zhao2018}]. This technique, called ``frequency-modulated capacitive tip-sample distance control mode'' \cite{Feng2022}, ensures that the tip remains lifted, eliminating the need for a first pass. \textcolor{reviewint}{However, these two methods require a highly stable environment and flat, electrically conducting samples -- conditions that are not always fulfilled, especially for patterned devices such as magnetic sensors or memory.}
    
    In this paper, we present a new method for achieving sensitive vacuum MFM measurements at high Q-factors while still enabling robust topography detection. This method is implemented in a \textit{Park Systems NX-Hivac Atomic Force Microscope}. We modify the commonly used two-pass mode to develop a high quality-factor two-pass mode. \textcolor{reviewint}{In the first pass, the cantilever's Q-factor is artificially lowered to allow for stable topography imaging. In the second pass, instead of measuring the phase shift with a lowered Q-factor, we measure the cantilever's oscillation frequency shift using a phase-locked loop (PLL). The PLL utilizes the maximum possible Q-factor, enabling the highest sensitivity in magnetic stray field measurements. Additionally, by using frequency shift and slope retracing in the second pass, non-linear phase response at high Q-factors and topographical interplay in the MFM signal are circumvented.}
    
    \textcolor{reviewint}{To explain the underlying physics, we introduce the concept of the Q-factor and methods for controlling it in the theory chapter. Significant focus is given to signal formation in atomic force microscopy, particularly in MFM. The theory section concludes with a brief discussion on noise in atomic force microscopy (AFM). In the experimental section, we describe the new high quality-factor two-pass mode. The improvement in sensitivity is demonstrated using a nano-patterned magnetic sample with etched topography. We validate the linearity of signal response through measurements of a Co/Pt multilayer film system forming stripe domains. To demonstrate topographical interplay and how to circumvent it, we use the nano-patterned magnetic sample again. Finally, we conclude that high quality-factor two-pass mode offers robust, reliable, linear, and stable measurements on samples with topographical features, making it widely applicable in commercial setups while significantly enhancing magnetic sensitivity.}

\section{\label{sec:theory}Theory}

    The two-pass mode (also referred to as \textit{lift mode} or \textit{interleave mode} by some manufacturers) is a well-established technique and is widely regarded as the workhorse of MFM \cite{Meyer2021}. Its fundamentals are covered extensively in various textbooks and articles on the subject \cite{Meyer2021, Voigtlaender2015, Haugstad2012, Kazakova2019, Winkler2023}. \textcolor{SS}{For the convenience of the reader, we comprehensively review the theoretical background of MFM here.} We begin by introducing the concept of the Q-factor. From this starting point, we establish the foundational theory governing the cantilever's oscillations and introduce the less commonly known Q-control operation \cite{Rodriguez2003, Hoelscher2007}.
    
    \textcolor{reviewint}{We then examine the frequency and phase responses of the cantilever's oscillation caused by tip-sample forces. Specifically, we will cover the phase- and frequency shifts that arise from the interaction between the magnetic tip and the sample's magnetic stray field in a dedicated section. Finally, a brief discussion of noise, which limits the smallest detectable signals in AFM, will conclude this theory overview.}

    \subsection{Quality-Factor in Dynamic AFM}
    
        The Q-factor, which quantifies the degree of damping in an oscillating system, plays a central role in determining the sensitivity of MFM measurements. It can be understood as the ratio of the energy stored in the oscillation to the energy dissipated per oscillation cycle \cite{Voigtlaender2015}.
    
\begin{figure}[tpb]
            \includegraphics[width=1\linewidth]{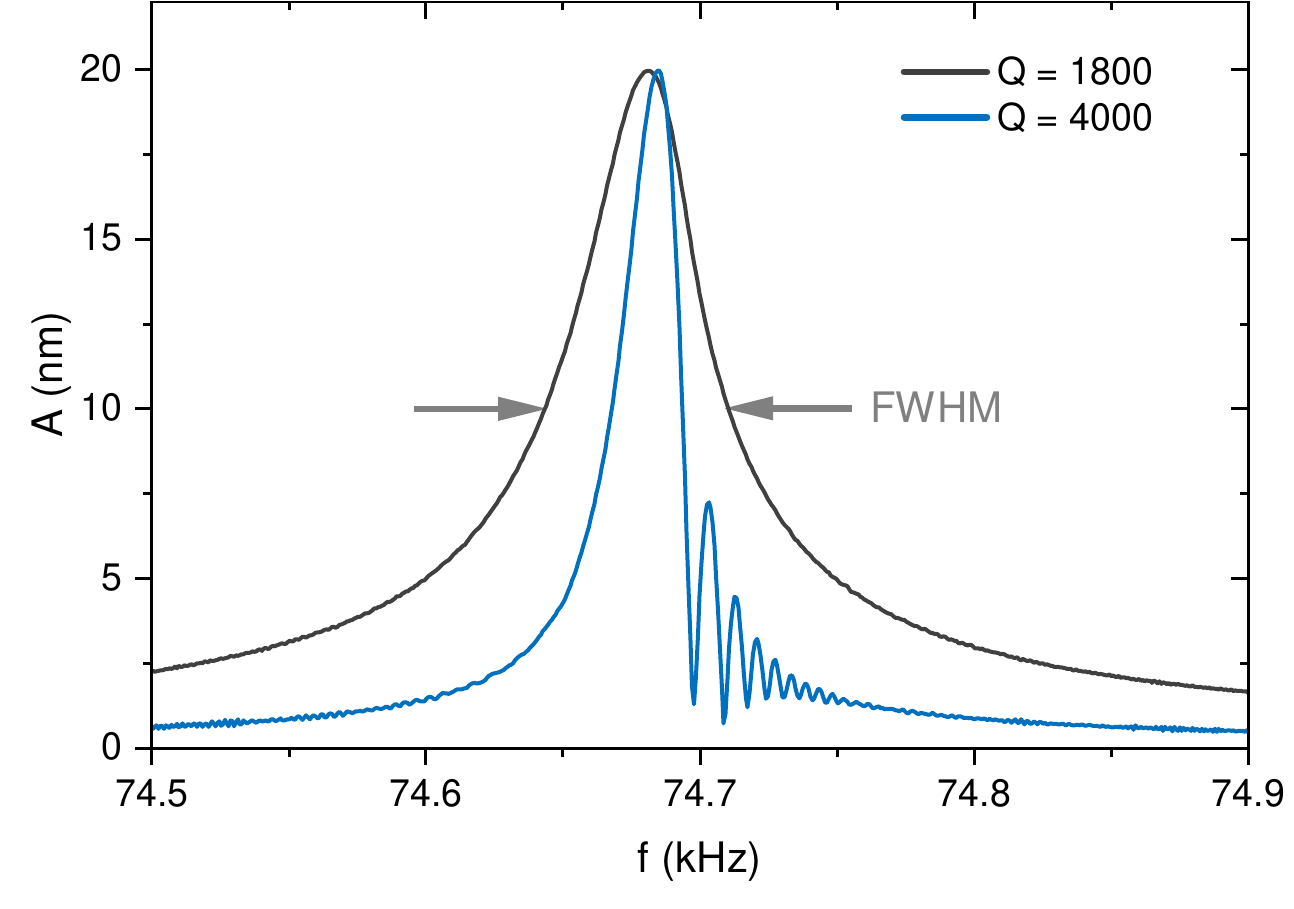}
            \caption{Ringing due to high Q-factor. Experimentally acquired frequency sweep data. The recorded amplitude A describes the peak-to-peak oscillation of the tip. Both curves were obtained with Q-control enabled, once with high Q-control attenuation (black curve, $Q \approx 1800$) and low attenuation (blue curve, $Q \approx 4000$). After reaching resonance at $f_0 = 74.68$\,kHz, ringing is observed in the blue curve, inhibiting operation.} 
            \label{fig:high-q-issues}
        \end{figure}

    
        In atomic force microscopy, vacuum conditions lead to higher Q-factors by reducing the density of gaseous particles, thereby minimizing collisions with the oscillating cantilever and effectively reducing friction. As a result, less energy is dissipated, leading to an increase in the Q-factor. For high Q-factors, an alternative description using the bandwidth definition is often employed:
    
        \begin{equation}
            Q = \frac{f_0}{\Delta f_\textrm{FWHM}} \label{eq:fwhm}
        \end{equation}
    
        where $f_0$ is the resonance frequency, and $\Delta f_\textrm{FWHM}$ is the full width at half maximum (FWHM) of the resonance peak. This method allows for easy determination of the Q-factor from non-contact frequency sweep data, as shown for $Q \approx 1800$ in Fig. \ref{fig:high-q-issues}. As the Q-factor increases, the resonance peak narrows, making the resonantly oscillating cantilever more sensitive to external forces, and thus increasing the overall sensitivity. Typical commercial MFM cantilevers exhibit Q-factors of around 200 in ambient conditions, but in vacuum, Q-factors can reach up to 20,000. Specially manufactured vacuum cantilevers can even achieve Q-factors as high as 200,000 \cite{Meyer2021}.
    
        For high Q-factors (greater than 2000), the oscillation becomes weakly damped, which makes stabilizing the amplitude in response to parameter variations more difficult. This is demonstrated in the frequency sweep for $Q \approx 4000$ shown in Fig. \ref{fig:high-q-issues}. Since only a small amount of energy is dissipated per cycle, transient effects, such as ringing or oscillatory transients, arise. These effects cause the cantilever to maintain its oscillation frequency even after the driving frequency has shifted, requiring a certain amount of settling time for the system to return to a steady-state harmonic oscillation.

        Q-factors can be artificially damped under vacuum conditions to mitigate this issue, using the Q-control mechanism discussed in Chap. \ref{sec:qcontrol}.

    \subsection{\textcolor{reviewint}{Cantilever Oscillation in AFM}} \label{sec:signal_gen_MFM}
        
        \textcolor{reviewint}{To introduce Q-control, it is essential to first understand the cantilever's oscillation.} In dynamic mode, the cantilever is excited at or near its resonance frequency. The motion $z(t)$ of the free cantilever (when it is not sensing a force) can be described by the equation of a driven harmonic oscillator:
        
        \begin{equation} \label{eq:diffeq}
            m \ddot{z}(t) + m \gamma \dot{z}(t) + k_\textrm{z} (z(t) - d)  = F_0 \cos(2\pi f_\textrm{d} t) 
        \end{equation}
        
        \textcolor{reviewer2}{where \(m\) is the mass, \(\gamma\) is the damping coefficient, \(k_\textrm{z}\) is the spring constant, \(d\) is the tip's equilibrium position, and the driving force \(F_0 = a_\textrm{d} k_\textrm{z}\) operates at driving amplitude \(a_\textrm{d}\) and driving frequency \(f_\textrm{d}\). The resonance frequency of the undisturbed oscillator is given by:}

        \begin{equation} \label{eq:res_freq}
        \textcolor{reviewer1}{
            f_0 = \frac{1}{2\pi} \sqrt{\frac{k_\textrm{z}}{m}} 
            }
        \end{equation}
        
        The damping factor \(\gamma\) can be described using the quality factor \(Q_0\) of the undisturbed oscillator, which interacts only with the surrounding gas. For frequencies close to the resonance frequency, i.e. \(f_\textrm{d} \approx f_0\), the damping factor is \(\gamma = 2\pi f_0 / Q_0\). With the ansatz \(z(t) = A \cos(2\pi f_\textrm{d} t + \varphi)\), the amplitude \(A\) and phase \(\varphi\) of the oscillator can be found as:
        
        \begin{eqnarray}
            A(f_\textrm{d}) &= \frac{F_0/(4\pi m)}{\sqrt{(f_0^2 - f_\textrm{d}^2)^2 + (f_0 f_\textrm{d} / Q_0)^2}} \\
            \varphi(f_\textrm{d}) &= \arctan\left( -\frac{f_0 f_\textrm{d}}{Q_0(f_0^2 - f_\textrm{d}^2)} \right) \label{eq:phi-tangens}
        \end{eqnarray}
        
        Key observations include that the amplitude \(A\) is primarily limited by the damping \(\gamma\) and \textcolor{reviewer1}{reaches its maximum at \(f_\textrm{max}^2 = f_0^2 \left( 1 - \frac{1}{2Q_0^2} \right)\), which approximates to \(f_\textrm{max} \approx f_0\) for large \(Q_0\).} Notably, the phase \(\varphi\) does not depend on the driving force, as this only affects the amplitude. A typical experimentally obtained curve of \(A\) and \(\varphi\) is shown in Fig. \ref{fig:phase-and-fm}, which will be discussed in a later section.

    \subsection{Q-control}\label{sec:qcontrol}

        To perform non-contact mode AFM measurements, external forces interacting with the oscillating tip must be considered. Additionally, an extra term is required to artificially reduce the Q-factor, which is essential for achieving Q-control. A more complete form of Eq. \ref{eq:diffeq} for AFM can be found in Ref. [\onlinecite{Hoelscher2007}]:
        
        \begin{eqnarray} \label{eq:diffeqexpand}
           &m \ddot{z}(t) + m \gamma \dot{z}(t) + k_\textrm{z} (z(t) - d) + \underbrace{g k_\textrm{z} z(t - t_0)}_{\textrm{Q-control}} \nonumber \\
           &= \underbrace{a_\textrm{d} k_\textrm{z} \cos(2\pi f_\textrm{d} t)}_{\textrm{external driving force}} + \underbrace{F_\textrm{ts}[z(t),\dot{z}(t)]}_{\textrm{tip-sample force}}
        \end{eqnarray}

        The first new term represents Q-control, characterized by the gain factor \(g\) and the time shift \(t - t_0\). \textcolor{reviewer2}{The second new term accounts for the tip-sample force \(F_\textrm{ts}\), which depends on both the tip's position \(z(t)\) and its velocity \(\dot{z}(t)\). Solving this equation analytically requires additional assumptions, as discussed in Ref. [\onlinecite{Hoelscher2007}]. For realistic tip-sample forces near the surface, non-linear effects dominate. However, steady-state solutions with sinusoidal cantilever oscillations are sufficient for measurement purposes. These solutions are derived by expanding \(F_\textrm{ts}\) into a Fourier series and neglecting higher-order terms. In the simplest case, when the cantilever oscillates far from the surface, the tip-sample force can be ignored (\(F_\textrm{ts} = 0\)), significantly simplifying the analysis.}
        
        One key result from Ref. [\onlinecite{Hoelscher2007}] is that Q-control allows the effective Q-factor, \(Q_\textrm{eff}\), to be adjusted by tuning the gain factor. Assuming \(F_\textrm{ts} = 0\) and \(f_\textrm{d} \approx f_0\), the effective Q-factor is given by:
        
        \begin{equation}
           Q_{\textrm{eff}}(g, t_0) = \frac{1}{1/Q_0 - g\sin(2\pi f_\textrm{d} t_0)}
        \end{equation}
        
\begin{figure}[tpb]
           \centering
           \includegraphics[width=0.49\textwidth]{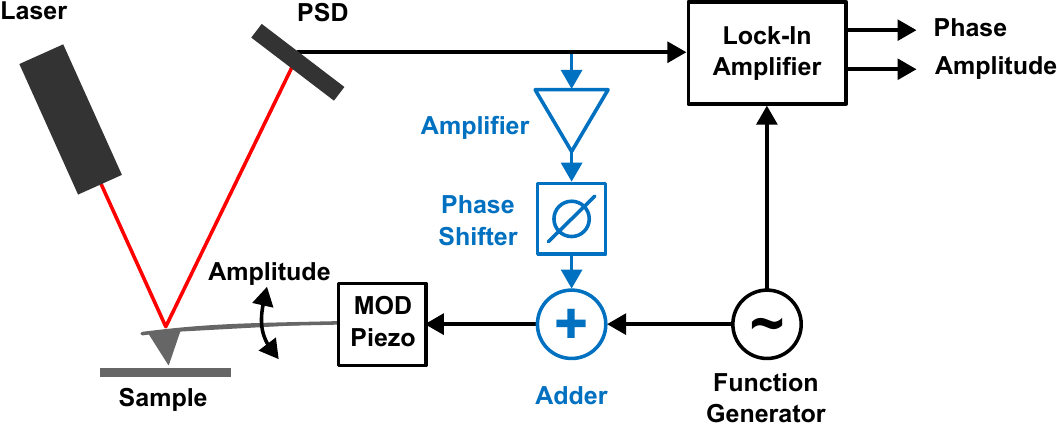}
           \caption{Q-control in AFM. The cantilever is excited via a function generator driving the modulation piezo (MOD Piezo). The induced oscillation is measured using the reflected laser signal on a position-sensitive photodiode (PSD). The electrical signal is fed to a lock-in amplifier, which measures phase and amplitude. By introducing a feedback path (blue) via an amplifier and phase shifter, the driving signal is modified, allowing Q-control.}
           \label{fig:qcontr-setup}
        \end{figure}

        The experimental realization of Q-control is illustrated in Fig. \ref{fig:qcontr-setup}. \textcolor{reviewint}{The cantilever's oscillation is induced by a function generator driving the modulation piezo (MOD Piezo), and the resulting oscillation is measured via a laser reflecting onto a position-sensitive photodiode (PSD). The signal is processed by a lock-in amplifier, which measures both the phase (relative to the function generator) and amplitude. By incorporating an additional feedback path with amplification and phase-shifting (e.g., time-shifting) electronics, the driving signal can be modified to compensate for or induce energy loss, thereby amplifying or attenuating the Q-factor.}

    \subsection{\textcolor{reviewint}{Frequency and Phase Shift in AFM}} \label{chap:non-linear}
        
        The second term on the right-hand side of Eq. \ref{eq:diffeqexpand} describes the influence of external forces. In the case of a force-free driven oscillator, the resonance frequency $f_0$ was introduced in Eq. \ref{eq:res_freq}. However, when an external force acts on the tip, it will shift the cantilever's resonance frequency. In typical cases, where \textcolor{reviewer1}{(i) the gradient of the tip-sample force $F_\textrm{ts}$ remains constant within the tip's oscillation amplitude,} (ii) the cantilever's restoring force behaves like a Hookean spring, $F_\textrm{cantilever} = -k_0 (s_\textrm{z} - s_{\textrm{z}_0})$, where $s_\textrm{z} - s_{\textrm{z}_0}$ represents the tip displacement around the equilibrium position $s_{\textrm{z}_0}$, and (iii) the restoring force is large \textcolor{reviewer1}{compared to the average tip-sample interaction force}, the impact of the force acting on the tip can be described as a modification of the spring constant \cite{Haugstad2012, Schroeter2009, Giessibl2001}:
        
        \begin{equation}
            k = k_0 + k' = k_0 - \frac{\partial F_\textrm{ts}}{\partial s_\textrm{z}} 
        \end{equation}
        
        Here, the differential expresses the \textcolor{reviewer1}{change in the} tip-sample force acting on the tip as it moves through the distance $s_\textrm{z}$ in the Z-direction during oscillation. Under this model, the resonance frequency is modified by the tip-sample force:
        
        \begin{equation}
            f_0' = \frac{1}{2\pi} \sqrt{\frac{ k_0 - \frac{\partial F_\textrm{ts}}{\partial s_\textrm{z}} \frac{k_0}{k_0}}{m}} = f_0\sqrt{ 1 - \frac{1}{k_0}\frac{\partial F_\textrm{ts}}{\partial s_\textrm{z}} } 
        \end{equation}
        
        Applying the Taylor expansion $\sqrt{1-x} \approx 1 - \frac{1}{2}x$, considering that the \textcolor{reviewer1}{derivative} of $F_\textrm{ts}$ is much smaller than the initial spring constant $k_0$, this approximation is justified and the modified resonance frequency $f_0'$ can therefore be approximated as:
        
        \begin{equation}
            f_0' \approx f_0 \left( 1 - \frac{1}{2k_0} \frac{\partial F_\textrm{ts}}{\partial s_\textrm{z}} \right) 
        \end{equation}
        
        Thus, the frequency shift is directly proportional to the change in the tip-sample force:
        
        \begin{equation}
            \textcolor{reviewer1}{
            \Delta f = f_0' - f_0 = - \frac{f_0}{2k_0} \frac{\partial F_\textrm{ts}}{\partial s_\textrm{z}} \label{eq:freq_shift} 
            }
        \end{equation}
        
        \textcolor{reviewer1}{Consequently, at a constant excitation frequency $f_\textrm{d}$, the observed phase shift $\varphi$ (as in Eq. \ref{eq:phi-tangens}) will change, since $f_0$ must be replaced by $f_0'$.} This is the fundamental working principle of MFM in two-pass mode, where this phase shift constitutes the measurement signal. By evaluating the first derivative of Eq. \ref{eq:phi-tangens},
        
        \begin{equation}
            \textcolor{reviewerboth}{
            \frac{\partial}{\partial f_\textrm{d}} \varphi(f_\textrm{d}) = - \frac{f_0 Q_0 (f_\textrm{d}^2 + f_0^2)}{Q_0^2 (f_\textrm{d}^2 - f_0^2)^2 + f_\textrm{d}^2 f_0^2}
            }
        \end{equation}
        
        it can be shown that for small $Q_0$, large \textcolor{reviewer1}{$f_\textrm{d}$}, and a limited variation of $f_0$ when \textcolor{reviewer1}{$f_\textrm{d} \approx f_0$}, it is reasonable to ignore the first term in the denominator, simplifying the equation to:
        
        \begin{equation}
            \textcolor{reviewerboth}{
            \frac{\partial}{\partial f_\textrm{d}} \varphi(f_\textrm{d} \approx f_0) = - \frac{2Q_0}{f_0} 
            }
        \end{equation}
        
        This results in a constant slope and, consequently, a linear signal response. This approximation is often adequate for MFM operation in air, with typical setups operating at $Q_0 \approx 200$ and $f_0 \approx 70$\,kHz. However, for large $Q_0$, this argument no longer holds, and non-linear behavior becomes significant in vacuum conditions.

    \subsection{\textcolor{reviewint}{Magnetic Forces and Signal Interpretation in MFM}}
        
        \textcolor{reviewer1}{In MFM, the force acting on the magnetically coated tip, with volume $V$ and local magnetization $\textbf{M}_\textrm{tip}(\textbf{r}',z')$ in the sample's stray field $\textbf{H}_\textrm{sample}(\textbf{r}',z')$, can be described as \cite{Schwenk2016, Meyer2021}:}
        
        \begin{eqnarray}  \label{eq:f_mag}
            \textbf{F}_\textrm{mag}(\textbf{r},z)=  &&\mu_0 \iint_{V'} \left( \vec\nabla \cdot \textbf{M}_\textrm{tip}(\textbf{r}',z')\right) \nonumber \\  
            &&\cdot \textbf{H}_\textrm{sample} (\textbf{r} + \textbf{r}', z+z') d\textbf{r}'dz'
        \end{eqnarray}
        
        where $\textbf{r} = (x,y)$ is the in-plane coordinate vector, $z$ is the measurement height, and $\mu_0$ is the vacuum permeability. \textcolor{reviewer1}{This equation can also be written in another form as \cite{Hug1998}:}
        
        \textcolor{reviewer1}{
        \begin{eqnarray}  \label{eq:f_mag_density}
            \textbf{F}_\textrm{mag}(\textbf{r},z) =  \mu_0 \int_{V'} \rho_\textrm{tip}(\textbf{r}',z') \textbf{H}_\textrm{sample} (\textbf{r} + \textbf{r}', z+z') dV' \nonumber \\   
            + \mu_0 \oint_{A'} \sigma_\textrm{tip} (\textbf{r}',z')  \textbf{H}_\textrm{sample} (\textbf{r} + \textbf{r}', z+z') dA'
        \end{eqnarray}
        }
        
        \textcolor{reviewer1}{where $\rho_\textrm{tip}(\textbf{r}',z') = \vec\nabla \cdot \textbf{M}_\textrm{tip}(\textbf{r}',z')$ is the tip's magnetic volume charge density, where $(\textbf{r}',z')$ are inside the tip volume $V$, and $\sigma_\textrm{tip}(\textbf{r}',z') = \textbf{M}_\textrm{tip}(\textbf{r}',z')\cdot \textbf{n}(\textbf{r}',z')$ represents the tip's magnetic surface charge distribution, where all $(\textbf{r}',z')$ are on the tip's surface $A$, and $\textbf{n}$ is the normal vector. Intuitively, Eq. \ref{eq:f_mag_density} expresses force as charge times field.}

        By inserting Eq. \ref{eq:f_mag} into Eq. \ref{eq:freq_shift}, the relation between the local magnetic field and the frequency shift of the oscillating cantilever can be derived. These calculations are conveniently performed in partial Fourier space with $(x,y,z) \rightarrow (k_x,k_y,z)$, as detailed in Refs. [\onlinecite{Hu2020, Hug1998, Schendel2000, Zhao2019, Schwenk2016}]. This results in:  
        
        \begin{eqnarray}
            \Delta f (\textbf{k},z) =  -\frac{\mu_0 f_0}{2k_\textrm{z}} \cdot \textrm{LCF}(\textbf{k},\theta,\phi, A_0) \nonumber \\
             \cdot \frac{\partial \textbf{H}_{\textrm{z,tip}}^*(\textbf{k},z)}{\partial_\textrm{z}}\cdot \textbf{H}_{\textrm{z}_\textrm{sample}}(\textbf{k},z) 
            \label{eq:freq_shift_MFM}
        \end{eqnarray}
        
        For \textcolor{reviewer1}{$f_\textrm{d} \approx f_0$}, and thus small $\Delta \varphi$, this gives:
        
        \begin{eqnarray} \label{eq:delta_varphi_MFM}
            \Delta \varphi (\textbf{k},z) = -\frac{\mu_0 Q}{k_\textrm{z}} \cdot \textrm{LCF}(\textbf{k},\theta,\phi, A_0) \nonumber \\
            \cdot \frac{\partial \textbf{H}_{z,\textrm{tip}}^*(\textbf{k},z)}{\partial_\textrm{z}}\cdot \textbf{H}_{\textrm{z}_\textrm{sample}}(\textbf{k},z) 
        \end{eqnarray}       
        
\begin{figure}[tpb]
            \centering
            \includegraphics[width=0.49\textwidth]{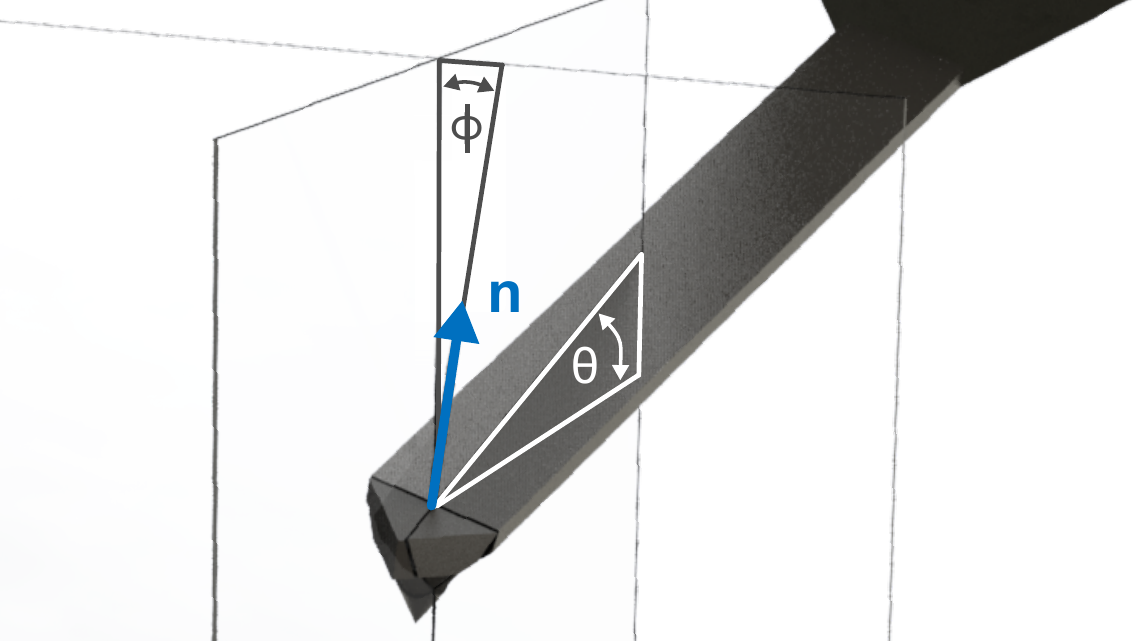}
            \caption{Canting angles due to cantilever tilt in MFM. Cantilevers are mounted at an angle $\theta$ to the sample surface, such that only the tip at the end of the cantilever interacts with the sample. Therefore, the normal vector $\textbf{n}$ of the cantilever (and hence the tip) is not aligned with the normal vector of the sample, a factor that must be considered for quantitative evaluation. Furthermore, the cantilever may be installed with a twist or the sample may not be mounted evenly, accounted for by the angle $\phi$.}
            \label{fig:Winkel_LCF_render}
        \end{figure}

        The introduced lever correction function $\textrm{LCF}$ accounts for cantilever- and device-specific parameters. It corrects for the canting angles $\theta$ and $\phi$ (see Fig. \ref{fig:Winkel_LCF_render}) and the finite oscillation amplitude $A_0$. The derivative of the complex conjugate of $\textbf{H}_{z,\textrm{tip}}$ describes the effective stray field gradient of the tip, located in a plane parallel to the sample surface at measurement height $z$. These terms convoluted with the stray field $\textbf{H}_{\textrm{z}_\textrm{sample}}$ of the sample yields the measured phase shift.
        
        Thus, damping the Q-factor in MFM phase shift measurements leads to a proportional reduction in the phase signal while introducing additional noise from the electronics, thereby further lowering the signal-to-noise ratio (SNR). While phase shift signal improvement is directly linked to the quality factor ($\varphi \propto Q$), in the case of frequency shift, this is not the case, as Eq. \ref{eq:freq_shift_MFM} is independent of $Q$. To understand the SNR improvement for frequency shift, a closer examination of the origin of noise in AFM is required.

    \subsection{\textcolor{reviewint}{Noise in AFM}}
        
        Thermal noise, arising from thermally induced cantilever motion in AFM, limits the minimum detectable force gradient as follows:
        
            \begin{equation} \label{eq:noise-afm}
            \textcolor{reviewer1}{
                \frac{\partial F_\textrm{min}}{\partial z} = \sqrt{ \frac{4k_0k_\textrm{B} TB}{2\pi f_0 Q \langle A_\textrm{osc}^2 \rangle}}
                }
            \end{equation}
        
        where $k_\textrm{B}$ is the Boltzmann constant, $T$ is the absolute temperature, $B$ is the bandwidth, and $\langle A_\textrm{osc}^2 \rangle$ represents the \textcolor{reviewer1}{mean square oscillation amplitude}. Depending on whether static or dynamic mode with amplitude or frequency modulation is used, a factor of $\sqrt{2}$ applies (further details in Refs.  [\onlinecite{Albrecht1991,Voigtlaender2015}]).
        
        This equation highlights the desirability of a high Q-factor to improve sensitivity. However, a high Q-factor also influences the required bandwidth in amplitude-modulated (AM) operation: if an external force acts on the cantilever, thereby altering the resonance frequency $f_0$, the oscillating system needs time to reach a new steady state. The time required for the system to respond is characterized by the time constant $\tau \approx 2Q/\omega_0 = Q/(\pi f_0)$. Consequently, for phase shift measurements, bandwidth and quality factor are interdependent, which can result in excessively slow measurements at high Q-factors. This limitation, however, does not apply to frequency shift measurements, where, by tracking $f_0$, the issue of settling time can be avoided. In this case, the bandwidth is limited only by the demodulation system used for frequency modulation (FM), rather than by transient behavior.

        \textcolor{reviewerboth}{Note that increasing the oscillation amplitude would theoretically improve the minimum detectable force gradient (see Eq.~\ref{eq:noise-afm}). However, for large amplitudes, the tip-sample force $F_\textrm{ts}$, averaged over one oscillation cycle, decreases, thereby reducing the signal. Additionally, very large oscillation amplitudes require a higher lift height, leading to a decay in the magnetic signal greater than the improvement in the noise floor. For quantitative evaluation, the derivative of $F_\textrm{ts}$ must also remain reasonably constant over the oscillation cycle. Therefore, the usable amplitude range is limited. Although, in theory, the elastic range of the cantilever allows oscillation amplitudes of 100\,nm or more, in practical MFM operation, amplitudes exceeding 20\,nm are not considered feasible.}

\section{\label{sec:experimental_setup}Experimental Setup}
    
    \textcolor{reviewint}{\textcolor{reviewer2}{In this section, the high quality-factor two-pass mode is introduced, which is a modified version of the commonly known two-pass mode.} A detailed explanation of signal formation in both the first and second pass of the two-pass mode is provided, culminating in the introduction of the high Q-factor two-pass mode.}
    
        \subsection{Phase and Frequency Detection}
    
        \subsubsection{First Pass: Topography}
            
\begin{figure}[tpb]
                \centering
                \includegraphics[width=0.49\textwidth]{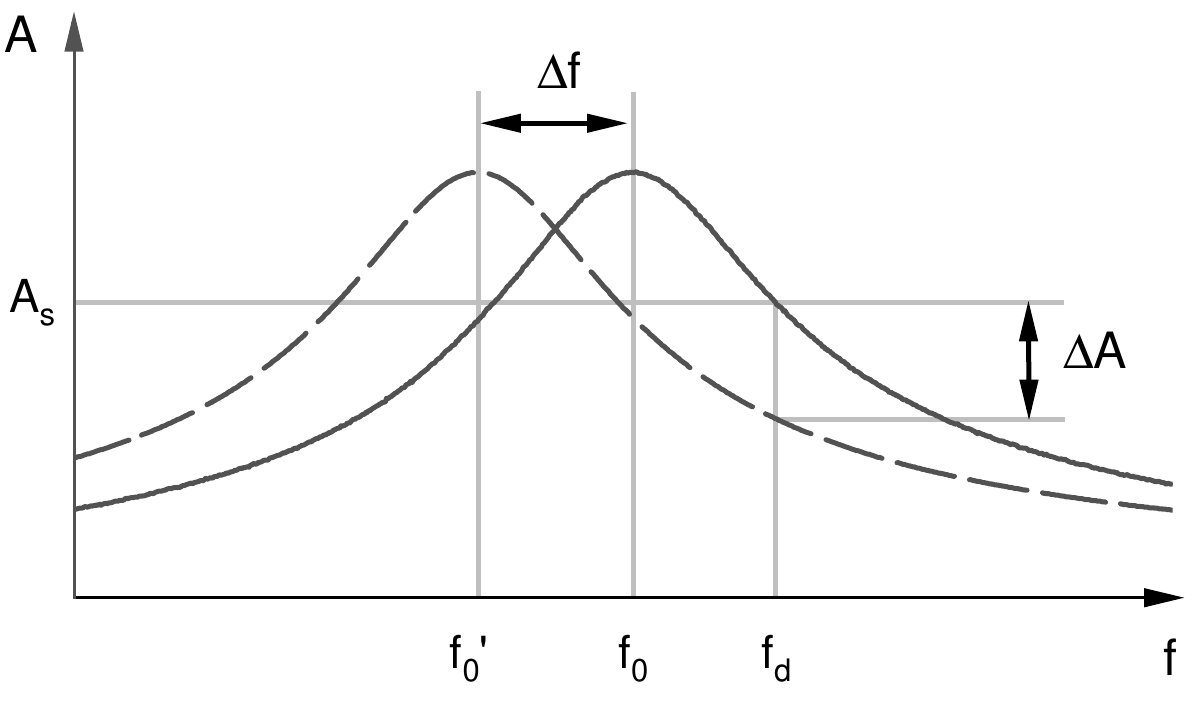}
                \caption{Topography acquisition via amplitude modulation. The freely oscillating cantilever shows a resonance curve around $f_0$ (solid line). When external forces act, the resonance frequency shifts, for instance, from $f_0$ to $f_0'$, causing the entire resonance curve to shift by $\Delta f$. The new steady state is indicated by the dashed line. The shift in resonance leads to an amplitude change $\Delta A$ at the fixed drive frequency $f_\textrm{d}$. This amplitude change is fed into a feedback loop, which adjusts the Z-piezo movement to maintain the amplitude at the setpoint $A_\textrm{s}$, allowing the sample's topography to be measured. \textcolor{reviewer1}{The curves have been simplified for explanation purposes.}}
                \label{fig:ampl-shift-operation}
            \end{figure}

            Fig.~\ref{fig:ampl-shift-operation} illustrates the working principle of amplitude-controlled topography measurements, as employed in the first pass of the two-pass mode. \textcolor{reviewer1}{The figure is simplified. In a real measurement, tip-sample forces distort the oscillation, affecting the resonance curve; see, for example, Ref.~[\onlinecite{Hoelscher2007}].} The freely oscillating cantilever exhibits a resonance peak, represented by the solid curve with a resonance frequency of $f_0$. For operation in non-contact mode, \textcolor{reviewer1}{the driving frequency $f_\textrm{d}$ must exceed $f_0$ \cite{Hoelscher2007}.} The driving frequency amplitude is chosen to achieve the desired oscillation  amplitude setpoint $A_\textrm{s}$. As the oscillating tip approaches the surface, external forces shift the resonance frequency, for example, from $f_0$ to $f_0'$, resulting in a frequency shift $\Delta f$ and a corresponding change in resonance behavior. This shift causes an amplitude change $\Delta A$ at the fixed drive frequency $f_\textrm{d}$, which is then used as feedback for the Z-piezo. The controller retracts or extends the Z-piezo to restore the setpoint amplitude $A_\textrm{s}$. The necessary piezo movement provides a map of the sample's topography. This method is effective for low Q-factors (e.g., $Q \approx 200$), where the resonance peak has a full width at half maximum (FWHM) of approximately 350 Hz, while the frequency shift is on the order of 10 Hz.
            
        \subsubsection{Second Pass: Magnetic Signal}
        
\begin{figure}[tpb]
                \centering
                \includegraphics[width=0.49\textwidth]{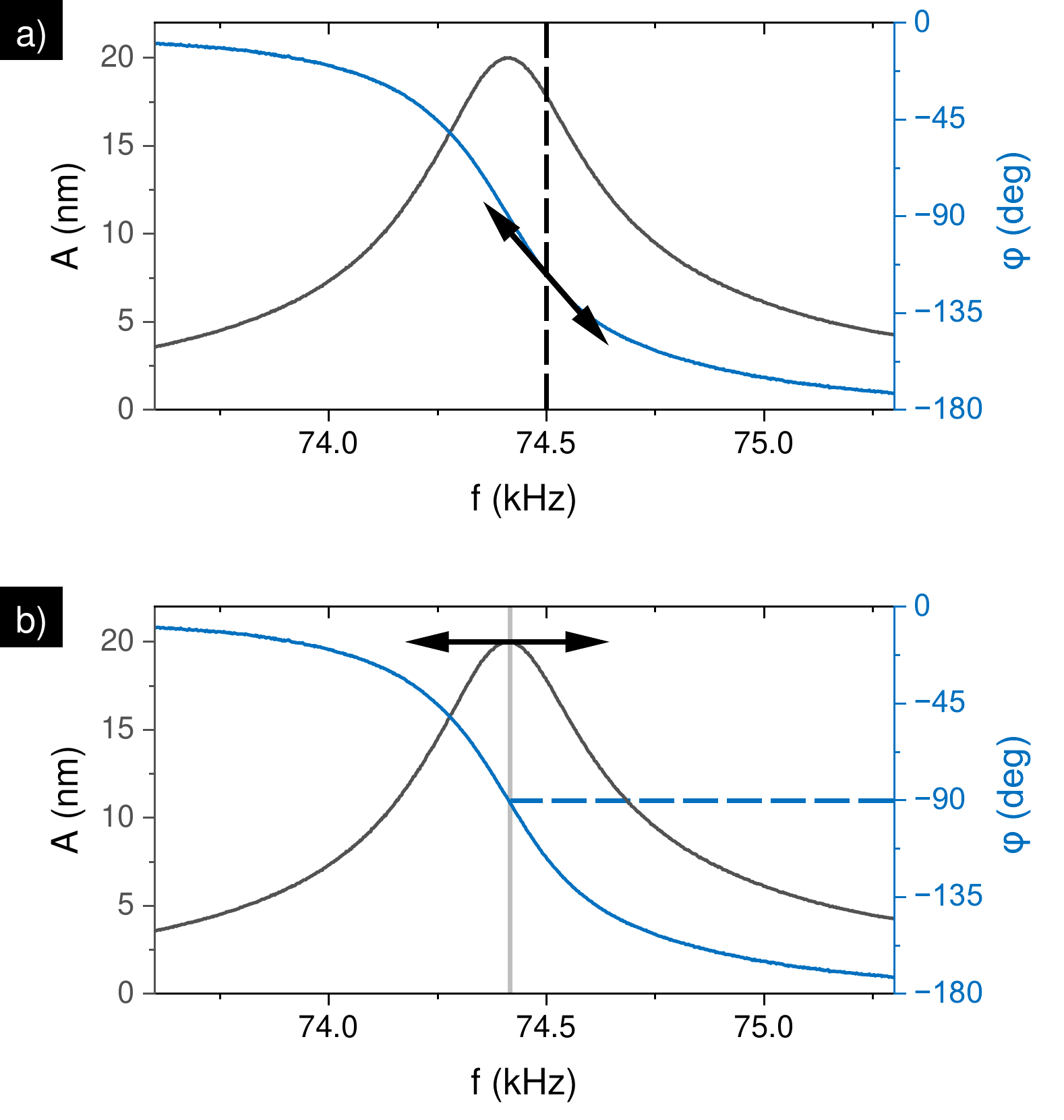}
                \caption{MFM operation with (a) phase and (b) frequency shift detection. In (a), phase shift detection is depicted as used in standard MFM two-pass operation. Operation occurs slightly off the resonance frequency: for tapping mode at lower frequencies, while for non-contact mode, as shown here, at higher frequencies. In lift mode, at the setpoint (vertical dotted line), the phase shift is measured. Near the resonance peak, the phase response behaves almost linearly (as indicated by the arrow). In (b), frequency shift detection is shown. The phase value at the resonance peak is used as the setpoint, enabling tracking of the peak and, consequently, the resonance frequency. Since the phase remains constant (controlled by a phase-locked loop), issues with non-linearity are avoided.}
                \label{fig:phase-and-fm}
            \end{figure}

            In the second pass (lift mode), the AFM controller retraces the topography acquired during the first pass, adding a user-defined lift height. The magnetic interaction between the tip and the sample induces a shift in the cantilever’s resonance frequency. In MFM, this magnetic interaction can be detected by either maintaining a constant excitation frequency and monitoring the phase shift or by tracking the changes in resonance frequency. Fig.~\ref{fig:phase-and-fm} illustrates both methods using experimentally obtained frequency sweep data for operation in air. The black curve represents the amplitude, while the blue curve shows the phase. The phase shift at resonance has been adjusted in post-processing to match -90 degrees.
            
            Under ambient conditions, phase shift detection is commonly used, as shown in Fig.~\ref{fig:phase-and-fm}a. For low Q-factors, the measured phase shift is relatively small, usually within single-digit degrees. Consequently, the phase response remains in a nearly linear range (as indicated by the arrow). 
            
            Under vacuum conditions, however, measurement signals can reach tens of degrees \footnote{The actual signal response depends on the tip and sample. ``Weak'' samples or tips with small magnetic stray fields may not exhibit such large responses, and their phase shifts may remain within single-digit degrees.}, pushing the response well beyond the linear range, which makes the data unsuitable for quantitative analysis. Therefore, in vacuum, frequency shift measurements are preferred, as they eliminate this issue. The resonance frequency (represented by the vertical line in Fig.~\ref{fig:phase-and-fm}b, which may shift in either direction) is tracked by setting the corresponding phase at resonance as the setpoint (in this case, -90 degrees, indicated by the horizontal line). A phase-locked loop (PLL) adjusts the excitation frequency, ensuring that the phase remains at the setpoint, thereby tracking the resonance frequency peak.

    \subsection{\label{sec:2passdual}\textcolor{reviewer2}{High Quality-Factor Two-Pass Mode}}
        
        The concept behind the new \textcolor{reviewer2}{high quality-factor two-pass mode} is to vary the Q-factor and signal detection scheme between the two passes. This approach optimizes stability during topography acquisition while enhancing sensitivity when measuring magnetic stray fields in lift mode. The measurement system used here is a Park Systems NX-Hivac AFM equipped with a signal extension module (SAM), which allows signal modification. The external lock-in amplifier (LIA) is a Zurich Instruments HF2LI equipped with a phase-locked loop (PLL). 
        
        \begin{figure}[tpb]
            \centering
            \includegraphics[width=0.49\textwidth]{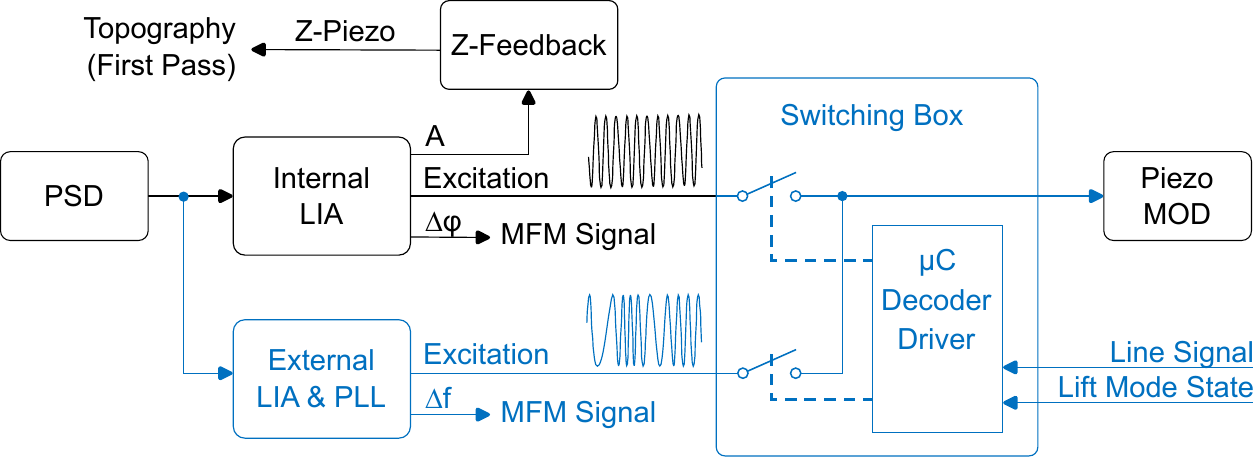}
            \caption{\textcolor{reviewer2}{Schematic overview of the new high quality-factor two-pass mode. The additional required setup is highlighted in blue, while the standard MFM setup is in black. During normal operation, the AFM’s internal lock-in amplifier (LIA) reads the position of the cantilever-reflected laser beam via a position-sensitive diode (PSD). The LIA provides the amplitude signal $A$, which is used by the Z-feedback circuit to control the Z-piezo, thereby obtaining the sample topography. In the second pass (lift mode), the phase-shift signal $\Delta\varphi$ serves as the MFM signal. For phase-locked loop (PLL) measurements, a second LIA is employed. In lift mode, the excitation (drive) signal applied to the modulation piezo (Piezo MOD) is switched from the internal LIA to the external LIA. While the internal LIA provides constant excitation at a set frequency, the external LIA tracks the resonance frequency shift via the PLL and adjusts the excitation accordingly. This frequency shift $\Delta f$ is used as the MFM signal. The corresponding excitation signal in the time domain is shown above the excitation signal path. Signal switching is achieved through a microcontroller (\textmu C) operated switching box, which connects the source of the excitation signal via CMOS analog multiplexers (combining decoder and driver) to the line signal and lift mode state signal from the microscope's scan generator.}}
            \label{fig:Dual-Mode_Paper}
        \end{figure}
        
        Figure~\ref{fig:Dual-Mode_Paper} provides a simplified overview of the new setup. In the first pass, the AFM's internal lock-in amplifier is used, utilizing Q-control to artificially lower the Q-factor to a manageable level for topography detection. \textcolor{reviewint}{It is important to note that topography is acquired in non-contact mode (NCM) to prevent damage to both the tip and sample, which is a prerequisite for quantitative magnetic force microscopy (qMFM) \cite{Hu2020, Hug1998, Schendel2000, Zhao2019, Schwenk2016, Sakar2021, Feng2022a}.}

        In lift mode, the AFM controls the lift height via the Z-piezo without requiring additional feedback. This enables the modulation piezo (Piezo MOD) to be switched and operated by the external HF2LI during lift mode at any time. Signal locking by the HF2LI occurs within a few hundred microseconds, allowing switching to take place during overscan (scanning a user-defined percentage beyond the desired scan area to prevent streaking at the image edges). The HF2LI excites and tracks the frequency of the oscillating tip via a phase-locked loop. The measured frequency shift from the PLL is transmitted to the microscope controller through an auxiliary input for data acquisition by the AFM measurement software.
        
        \begin{figure*}[tpb]
            \textcolor{reviewer2}{
            \centering
            \includegraphics[width=0.99\textwidth]{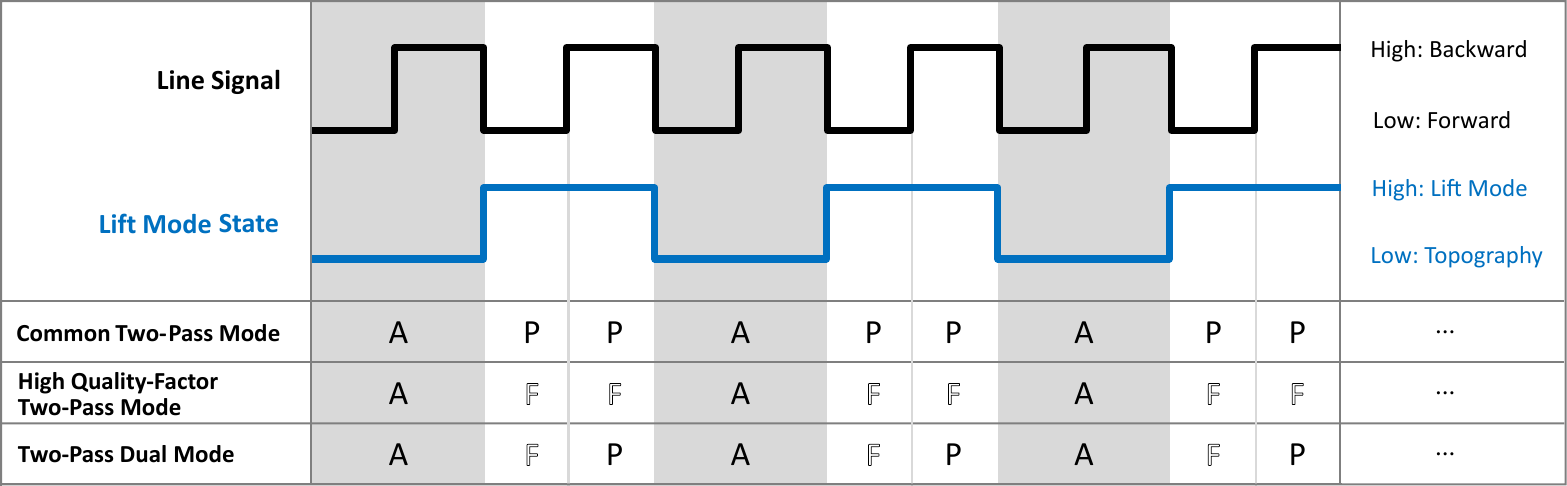}
            \caption{Timing diagram illustrating the modifications to the two-pass mode. The line signal and lift mode state are input into the microcontroller. Depending on the operation mode, the drive signal applied to the modulation piezo is switched between the external lock-in amplifier (which runs a phase-locked loop) and the internal lock-in amplifier. Topography is always acquired using the internal lock-in amplifier, which operates at a dampened low Q-factor via amplitude modulation (marked A in the table). In the common two-pass mode, the phase shift (marked P) is measured during the lifted pass, again at low Q-factor. In contrast, in the high quality-factor two-pass mode, the frequency shift (marked F) is measured during the lifted pass using a phase-locked loop at the maximum possible Q-factor with the external lock-in amplifier. For comparison measurements, the two-pass dual mode is employed, where the frequency shift is measured at high Q-factor in the forward direction and the phase shift at low Q-factor in the backward direction, enabling direct comparison.}
            \label{fig:timing-dual-mode}
            }
        \end{figure*}
        
        Signal switching is achieved through a custom-built switching box that uses a microcontroller (µC) to control several DG409 CMOS analog multiplexers (each equipped with a decoder and driver) interconnecting the AFM with the external lock-in amplifier. The line signal, which indicates scan direction (forward or backward), and the lift mode state are fed into the µC. Based on these inputs and the selected operation mode, the µC directs the excitation signal either to the AFM's internal lock-in amplifier with Q-control or to the external HF2LI to drive the modulation piezo. A graphical user interface (GUI) allows users to adjust the µC operation and select from various operating modes. Using the discussed techniques, three distinct measurement modes, as shown in Fig.\ \ref{fig:timing-dual-mode}, can be made available:

        \begin{itemize}
            \item \textbf{Common Two-Pass Mode.} The standard and widely used two-pass mode, where no signal switching occurs. Since the Q-factor is dampened (reducing sensitivity), this mode is typically used for an initial overview.
            \item \textcolor{reviewer2}{\textbf{High Quality-Factor Two-Pass Mode.} In this mode, the frequency shift is measured during lift mode in both forward and backward direction. Every time lift mode is entered, the signal is switched to the external HF2LI lock-in amplifier. This mode replaces the standard two-pass mode.   }   
            \item \textcolor{reviewer2}{\textbf{Two-Pass Dual-Mode.} To allow for direct comparison of phase and frequency shift measurements without recording two independent images, we introduce further the two-pass dual-mode. In this mode, the frequency shift is measured in the forward direction, while in the backward direction, the phase shift is measured using Q-control. This approach operates as quickly as the normal two-pass mode but eliminates the redundant control trace, which can otherwise help less experienced users identify problematic measurement settings, such as inappropriate scan speed that can cause discrepancies between forward and backward data.}
        \end{itemize}

\section{Experimental Results}
    
    \textcolor{reviewint}{In this section, the experimental results are presented. By measuring the frequency shift (instead of phase shift) in the second pass, the highest possible Q-factor can be utilized without losing sensitivity or encountering non-linearity in the measurement signals. The signal improvement between phase and frequency detection is demonstrated by measuring a nano-patterned magnetic sample with topographical features. In particular, the critical issue of non-linear phase response at high Q-factors is circumvented by measuring the frequency shift, as demonstrated by the measurement of the stray field of a Co/Pt multilayer system forming symmetric stripe domains. Finally, the interplay of topography in standard two-pass lift mode -- which follows the sample’s topography -- is examined by comparing it with a measurement retracing sample's slope in the second pass.}

    \subsection{\textcolor{reviewint}{Improvement of Signal-to-Noise Ratio}}
        
\begin{figure}[tpb]
            \centering
            \includegraphics[width=0.49\textwidth]{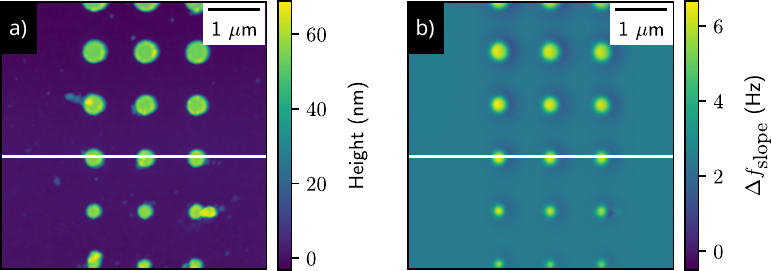}
            \caption{Nano-patterned magnetic sample. (a) Topography overview. The three circles with a diameter $d=300$\,nm (marked by the white line) have been used for the line profiles that are evaluated. All structures are 60\,nm high. Multilayer system: Ta(5)/Pt(8)/[Co(1)/Ru(1.4)/Pt(0.6)]$_{10}$/Pt(2.4). (b) Corresponding MFM frequency shift image acquired in vacuum at $Q \approx 9000$.}
            \label{fig:af_sf6_2d}
        \end{figure}

        The feasibility of the new high quality-factor two-pass mode is demonstrated by measuring a nano-patterned magnetic sample that combines topographical features with low magnetic stray fields. The sample consists of circles of varying sizes, which establishes a dot array. For evaluation, three circles with a diameter of $d=300$\,nm and a height of $h=60$\,nm were selected (see Fig.~\ref{fig:af_sf6_2d}a for the sample topography). The sample is composed of a \textcolor{reviewer1}{Ta(5)/Pt(8)/[Co(1)/Ru(1.4)/Pt(0.6)]$_{10}$/Pt(2.4)} multilayer stack (with layer thicknesses in nm) on Si, featuring perpendicular magnetic anisotropy. More details can be found in [\onlinecite{FernandezScarioni2021}]. Since the measurements are performed in two-pass dual-mode, it is ensured that phase and frequency shifts are measured in immediate succession, allowing direct comparison between the phase and frequency measurements. A full MFM image obtained by frequency shift measurement in vacuum at a Q-factor of $Q \approx 9000$ is shown in Fig.~\ref{fig:af_sf6_2d}b. As the sample features structures with 60\,nm topography, the follow-slope mode was used to eliminate topography interference in the MFM signal (see Chap.~\ref{chap:follow-slope}).

        \begin{figure}[tpb]
            \centering
            \includegraphics[width=0.49\textwidth]{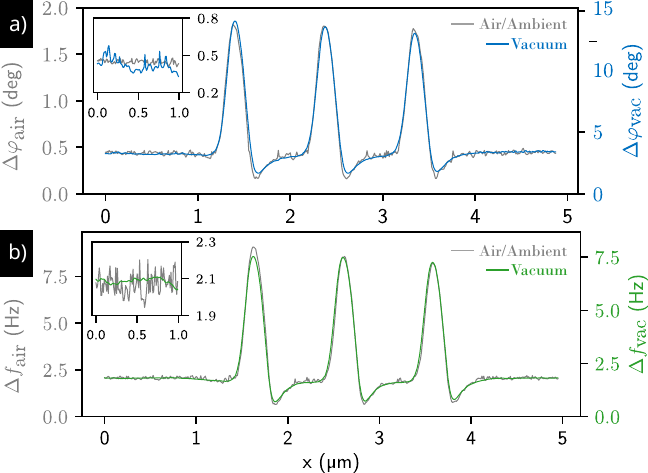}
            \caption{Line profile of the structured magnetic sample in air and vacuum, utilizing both phase shift and frequency shift signal detection. (a) Phase shift data: The gray profile represents a measurement in air with phase shift detection for $Q \approx 170$, exhibiting clear noise contribution. The blue profile corresponds to a phase shift measurement in vacuum with a dampened $Q \approx 1600$. Note the different scaling on the y-axes. The inset in the top left corner provides a zoomed-in view of a small excerpt of the signal, using identical y-axis scaling. (b) Frequency shift data: The gray profile indicates a measurement in air detecting frequency shift with $Q \approx 170$, showing dominant noise contribution. The green profile represents a vacuum measurement with frequency shift detection at $Q \approx 9000$. Similar y-axis scaling is applied, \textcolor{reviewer2}{deviation is likely caused by a variation in lift-height.} The inset in the top left corner also zooms into a small excerpt of the signal, maintaining identical y-axis scaling. All measurements were acquired with a scan rate of 0.05\,Hz. Calibration of the piezo scanners in air differs from that in vacuum; therefore, the x-axis has been rescaled to match the calibration. This issue can be circumvented by creating a vacuum calibration file for operation in vacuum.}
            \label{fig:vergleich_nur_linien}
            \label{fig:paper-af_sf6_v}
        \end{figure}
        
        The results of measurements for different Q-factors are depicted in Fig.~\ref{fig:vergleich_nur_linien}. The four line plots show:  In Fig.~\ref{fig:vergleich_nur_linien}a the phase shift signal in air (gray line profile, $Q \approx 170$) and phase shift signal in vacuum (blue line profile, $Q \approx 1600$); and in Fig.~\ref{fig:vergleich_nur_linien}b the frequency shift signal in air (gray line profile, $Q \approx 170$) and the frequency shift signal in vacuum (green line profile, $Q \approx 9000$). \textcolor{reviewint}{The slight asymmetry in the dips is due to the canting angle, which has not been corrected in this measurement to maintain a comparison of the raw data. Canting angles can be corrected using qMFM routines, which would also filter noise. The scaling on the y-axis in both  Fig.~\ref{fig:vergleich_nur_linien}a and  Fig.~\ref{fig:vergleich_nur_linien}b has been adjusted so that the signal traces of air and vacuum measurements coincide. To illustrate the origin of the signal-to-noise ratio (SNR) improvement, an excerpt of both traces with identical y-scaling is shown via insets in the upper left corners.}
        
\begin{figure}[tpb]
            \centering
            \includegraphics[width=0.49\textwidth]{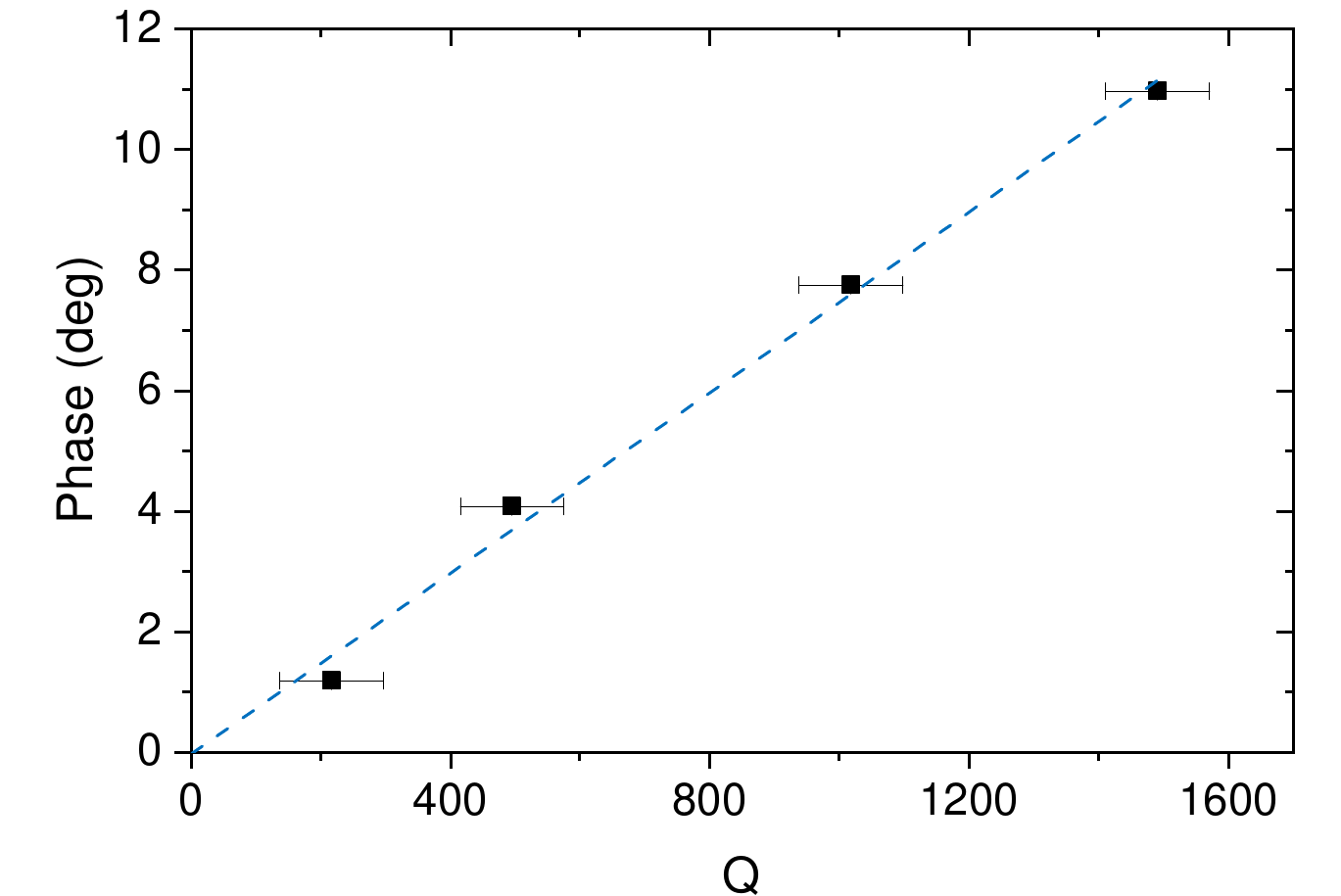}
            \caption{Peak-to-peak phase signal for increasing Q-factors. For small phase signals, the improvement behaves linearly with the Q-factor. A slope of $\Delta\varphi = 0.74$~deg per $\Delta Q = 100$ is observed. Note that at large Q-factors, non-linear effects come into play, causing deviations for large phase shift values (not shown here). Uncertainty bars account for drift, \textcolor{reviewer1}{with the worst Q-factor drift during the measurement series being 80.}}
            \label{fig:q-vs-phase}
        \end{figure}

        As expected, both detection modes show improved SNR when operating in vacuum compared to ambient conditions. As discussed in the theoretical section, the improvement in phase shift measurements stems from an increase in the absolute phase shift signal, while the noise floor remains unchanged. This observation is confirmed by the experiment: \textcolor{reviewer1}{the overall noise remains within the same order of magnitude. We attribute the higher noise levels in air to acoustic noise coupling, as the vacuum chamber lacks an acoustic enclosure for ambient operation.} The improvement of the phase signal scales linearly with the Q-factor (see Table~\ref{tab:table-snr} and Fig.~\ref{fig:q-vs-phase}), with the phase signal increasing by \textcolor{reviewer1}{$\Delta\varphi = 0.74$~deg} for every $\Delta Q = 100$. \textcolor{SS}{However, this linear relationship holds only for small values of $\varphi$, as will be demonstrated in the next section.} 
        
        Additionally, the noise contribution from Q-control increases with rising Q, eventually offsetting the improvement in phase signal, as reflected in the SNR values in Table~\ref{tab:table-snr}. Based on these results, we conclude that for this specific setup and batch of cantilevers, the optimal Q-factor for phase measurements is around $Q \approx 1000$.
        
        \textcolor{reviewer1}{Since Q is subject to (thermal) drift, we account for a general uncertainty of $\pm 80$, which represents the worst observed offset during Q-control operation. As $\varphi \propto Q$ (see Eq.~\ref{eq:delta_varphi_MFM}), this uncertainty, along with other factors, propagates to the measured peak-to-peak signal values and, consequently, the SNR in Table~\ref{tab:table-snr}. No statistical analysis has been conducted; therefore, no evaluation following Ref.~[\onlinecite{BIPM2008}] has been performed.} 

        \begin{table}[tpb]
            \begin{ruledtabular}
                \begin{tabular}{lrlllc}
                                &       & Q       & RMS Noise       & Signal        & SNR     \\
                \midrule 
                     Frequency  & Air   & 217     &  97.2 mHz    & 5.18 Hz        & 53.2 \\ 
                                & Vac   & 9117    &  10.8 mHz   &  5.52 Hz        & 512 \\ 
                     Phase      & Air   & 217     &  27.8 mdeg  & 1.19 deg     & 42.6 \\ 
                                & Vac    & 495     &  14.5 mdeg  & 4.08 deg     & 281 \\ 
                                & Vac    & 1018    &  17.3 mdeg  & 7.75 deg     & 448 \\ 
                                & Vac    & 1491    &  39.6 mdeg  & 11.0 deg     & 277 
                \end{tabular}
            \end{ruledtabular}
            \caption{Comparison of phase and frequency shift measurements in air and vacuum. The Q-factor is calculated from frequency sweep data. The root mean square (RMS) noise is obtained from scans conducted far away from the magnetic surface (at a lift height of 100\,µm, i.e., in stray-field-free space). The signal represents the peak-to-peak values from the same area of interest (the structured magnetic circles, as shown in Fig.~\ref{fig:paper-af_sf6_v}). The signal-to-noise ratio (SNR) is calculated based on these values. While the phase signal improves linearly with increasing Q-factor, the noise contribution from Q-control negates this improvement for this specific cantilever at Q-factors around 1000. \textcolor{reviewer1}{The poorer noise performance in air is most likely due to acoustic noise coupling, as no acoustic enclosure is available for non-vacuum operation.}}
            \label{tab:table-snr}
        \end{table}
            
        \textcolor{reviewer2}{For the frequency shift measurement, the signal amplitude remains relatively constant, while the noise decreases significantly, resulting in an improved SNR. According to theory, the peak-to-peak signal should remain unchanged; however, the measurement shows an offset of a few percent. We attribute this to the follow-slope measurement mode, which will be described in Section \ref{chap:follow-slope}. In the presence of topography, the slope experiences an offset, causing a variation in lift height by a few nanometers. Measuring at a different lift height affects the observed frequency shift. Given that the structures are relatively large, we expect that the decay of $\mathbf{k}$ in Fourier space is not significant, affecting only the measured frequency shift and not the overall signal shape.}
        
        \textcolor{SS}{In summary, we have demonstrated that higher Q-factors lead to a clear improvement in the SNR of the measurement data in both detection modes. For phase shift measurements, this improvement can be mainly attributed to an increase in the signal, while for frequency shift measurements, the noise level decreases as the Q-factor increases. While phase shift measurements require artificially lowering the Q-factor for stable operation, frequency shift measurements are limited only by the maximum Q-factor the cantilever can provide \footnote{Since an air-class cantilever was used for this comparison, the observed Q-factor of approximately 10\,k is far below the upper limit of 200\,k, achievable with carefully manufactured high Q-factor vacuum cantilevers.}. Another key advantage of frequency shift measurements is the elimination of non-linear behavior, as will be demonstrated in the following section.}

        \subsection{\textcolor{reviewint}{Overcoming Non-Linear Phase Response via Frequency Shift Measurement}}
            
\begin{figure*}[tpb]
                \centering
                \includegraphics[width=0.99\textwidth]{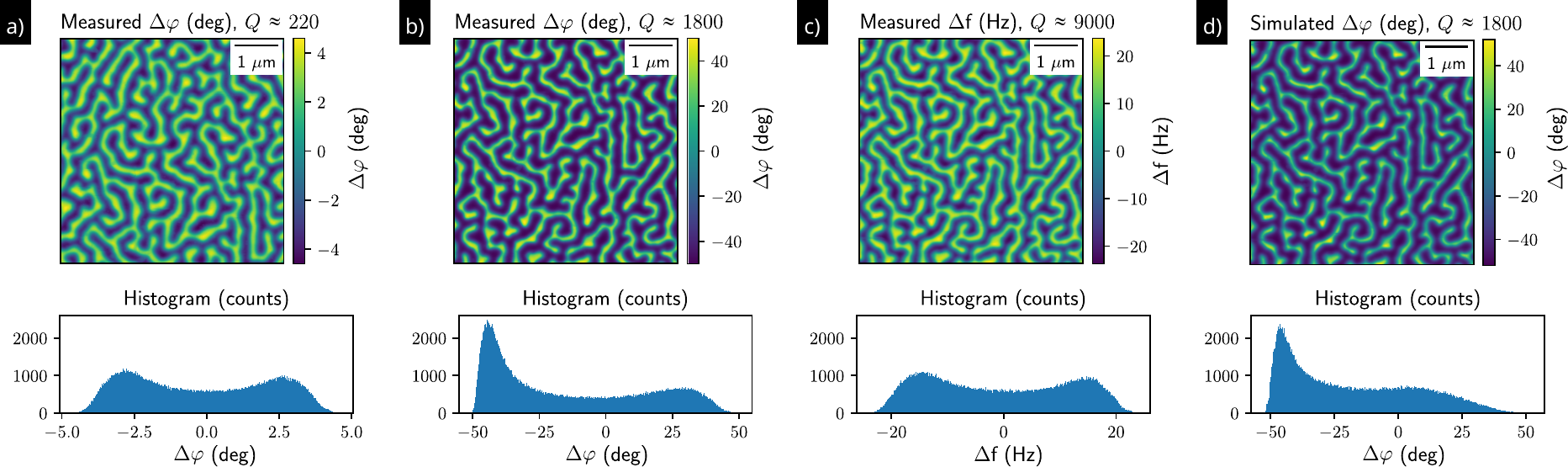}
                \caption{Measurement of a qMFM reference sample for different Q-factors. The well-known, calculable Co/Pt multilayer sample with the layer architecture Pt(2\,nm)/[(Co(0.4\,nm)/Pt(0.9\,nm)]$_{100}$/Pt(5\,nm)/Ta(5\,nm) on Si is used in this comparison measurement. The sample has been characterized in detail in [\onlinecite{Hu2020}]. While in (a) the domain distribution is even (equal amounts of bright and dark domains, with mirror symmetry in the histogram) in ambient conditions for low Q-factors, in (b) under vacuum conditions with a damped Q-factor ($Q \approx 1800$) and phase shift detection, the dark domains become more pronounced, as shown by the asymmetry in the histogram. In (c), the corresponding frequency shift image is plotted (two-pass dual-mode, $Q \approx 9000$). Since the two-pass dual-mode was used (meaning the sample could not have changed its physical properties in between), this observation cannot be explained by tip-sample interaction or other effects and must originate from the measurement principle itself. To further illustrate this, the data in (c) has been processed through Eq.~\ref{eq:phi-tangens}, yielding image (d). A similar asymmetry (as seen in (b)) with dominant dark domains is observed. \textcolor{reviewer1}{Note: Due to the phase response as portrayed in Fig.~\ref{fig:phase-and-fm}, the arccot function must be used instead of arctan when calculating (d).}}
                \label{fig:paper-domains-q}
            \end{figure*}

            The origin of non-linear behavior has been extensively discussed previously. Here, this effect is demonstrated on a well-known, calculable multilayer reference sample \cite{Hu2020} that forms up and down magnetized domains in a maze pattern, resulting in equal areal percentages of bright and dark areas (see Fig.~\ref{fig:paper-domains-q}a). However, in phase shift measurements with increasing Q-factors, a higher areal percentage of dark domains can be observed, \textcolor{reviewer1}{while the bright domains are suppressed}, as shown in Fig.~\ref{fig:paper-domains-q}b. For convenience, all images are accompanied by their corresponding histograms. In Fig.~\ref{fig:paper-domains-q}a, the domain pattern was measured under ambient conditions ($Q \approx 220$) using phase shift, showing an equal domain distribution. In Fig.~\ref{fig:paper-domains-q}b, the same reference sample was measured in vacuum ($Q \approx 1800$) using phase shift and Q-control. The Q-factor was chosen as large as possible to effectively illustrate the effect. The dark domains are much more pronounced, which is clearly visible in the histogram. Without prior knowledge of the phase behavior, this could easily be misinterpreted as tip-sample interaction, a sample defect, or, worse, as real measurement data. This could pose a significant risk when interpreting data for material characterization and quantitative measurements. However, with this setup, we can rule out these effects as the cause, since -- again -- the two-pass dual-mode was used, which acquires phase shift and frequency shift simultaneously. Consequently,  Fig.~\ref{fig:paper-domains-q}c shows the exact same position of the sample with the same AFM measurement parameters as Fig.~\ref{fig:paper-domains-q}b, but now with the modulation piezo driven by the external (HF2LI) lock-in amplifier that is measuring frequency shift.   The frequency shift data shows equally distributed dark and bright domains, indicating that the imbalance in the phase shift distribution originates solely from the measurement technique.

            The origin of the observed asymmetry in the domain distribution can be explained directly by Eq.~\ref{eq:phi-tangens} and the corresponding phase curve in Fig.~\ref{fig:phase-and-fm}. As the setpoint is slightly off-peak, the arctan is slightly off point symmetry, causing positive phase shift values to enter the non-linear regime more rapidly, resulting in a reduced signal. This not only decreases the absolute values but also breaks the symmetry of the corresponding peaks (as clearly observed in the histogram). By applying Eq.~\ref{eq:phi-tangens} (with $f_0 = 74660$\,Hz and an actual operation frequency 30\,Hz above $f_0$) to the frequency values of Fig.~\ref{fig:paper-domains-q}c, the corresponding Fig.~\ref{fig:paper-domains-q}d can be calculated, which aligns well with the measured data in Fig.~\ref{fig:paper-domains-q}b.
            
            Conversely, if the corresponding phase curvature is acquired in advance, these non-linearities could be compensated for in post-processing by adjusting the measured phase values to those expected if they were acquired linearly. However, as the arctan loses slope when far from the point of symmetry, sensitivity is reduced. Correcting these values would amplify noise to the point where no signal can be recovered, which is highly undesirable. This highlights the usefulness of the high quality-factor two-pass mode, which avoids these issues.

    \subsection{\label{chap:follow-slope}Topography Interplay}

        \begin{figure}[tpb]
            \centering
            \includegraphics[width=0.99\linewidth]{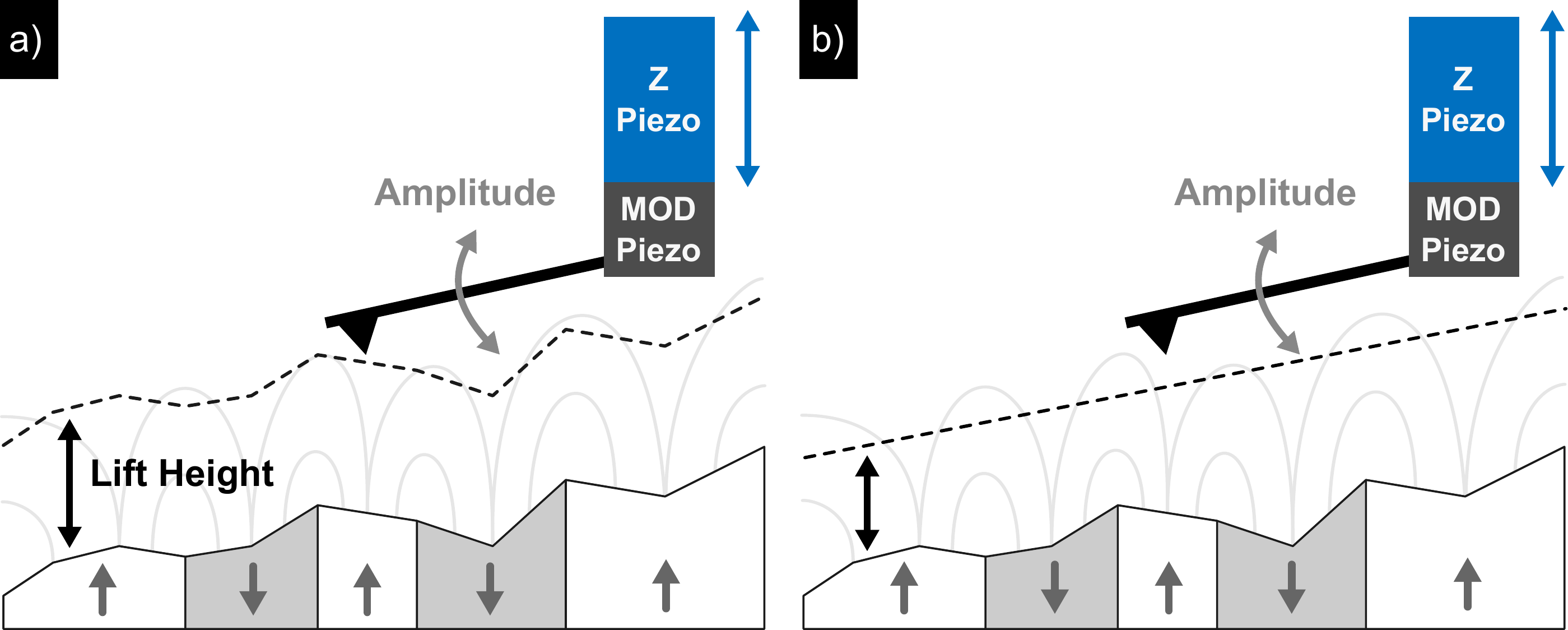}
            \caption{MFM operation in two-pass (lift) mode. (a) Second pass with lifted cantilever. The topography from the first pass is retraced. While the short-ranged Lennard-Jones potential (that is used in the first pass to map the surface) cannot influence the cantilever oscillation at this height, the long-ranged magnetic stray fields interact with the magnetic tip. (b) Second pass that does not retrace the topography but only the sample tilt (slope). This is useful for flat samples or samples with known structure to eliminate topography interplay in the MFM signal.}
            \label{fig:two-pass-mode}
        \end{figure}
        
        \textcolor{reviewint}{As we have demonstrated, the high quality-factor two-pass mode allows measurements without the observed non-linearities of the phase shift signal seen in classical two-pass mode. However, we note that other measurement artifacts may introduce additional deviations that significantly alter the MFM signal.} 

        One example we discuss here is the influence of topography. In two-pass mode measurements, the surface is typically retraced in the second pass (see Fig.~\ref{fig:two-pass-mode}a). However, following the topography in lift mode influences the detected magnetic signal. For instance, non-magnetic dirt on a flat magnetic sample could be mistaken for a magnetic signal, as the dirt will cause an additional lift height in the second pass, moving the cantilever out of the sample’s stray field and thus altering the magnetic signal.
        
\begin{figure}[tpb]
            \centering
            \includegraphics[width=0.49\textwidth]{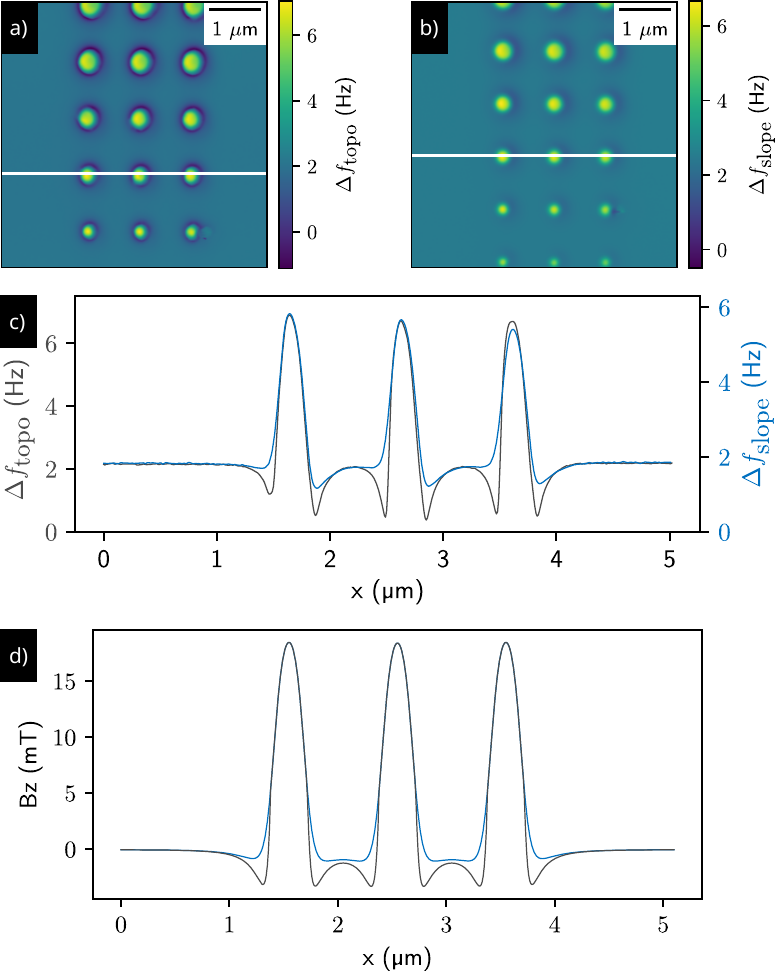}
            \caption{MFM measurements in common lift mode and follow slope mode. Both measurements were acquired in two-pass dual-mode, measuring frequency shift at a Q-factor of $Q \approx 9000$. (a) Commonly used lift mode operation that retraces the sample's topography in the lifted pass. The circular structures show a dark shadow around their edges. (b) Same measurement performed in follow slope mode. The dark shadows decreased while the overall peaks remained constant. (c) Line profiles along the white line in the scans for better visualization of the effect at hand. Gray: Line profile corresponding to (a). Blue: Line profile corresponding to (b). The dips emerging due to the topography retrace are noticeably less pronounced and are solely a measurement artifact. The slight asymmetry in the dips is due to the canting angle, which has not been corrected in this measurement, as here raw data is shown and not a qMFM measurement. (d) Simulated magnetic stray field in the z-direction ($B_z$) for a constant follow slope lift height of 120\,nm (blue) and topography-tracing height of 60\,nm (black). Good overall agreement with the effect observed in (c) is evident.}
            \label{fig:slope-vs-topo} 
        \end{figure}

        The issue of topography interplay becomes particularly problematic when working with artificially nano-patterned samples, as demonstrated in Fig.~\ref{fig:slope-vs-topo}. Here, we apply the high quality-factor two-pass mode to measure the three circles examined in Figs.~\ref{fig:af_sf6_2d} and \ref{fig:paper-af_sf6_v} -- but now in the common two-pass lift mode that retraces topography. When the topography of the circular structures is followed in lift mode, the frequency shift shows dark shadows around the dots, as shown in Fig.~\ref{fig:slope-vs-topo}a and the corresponding gray-colored line profile in Fig.~\ref{fig:slope-vs-topo}c.
        
        To avoid these artifacts, we use a mode that follows the average slope of the topography, as shown in Fig.~\ref{fig:two-pass-mode}b. In this mode, a linear slope is fitted through the measured topography, allowing the sample’s tilt to be traced in the second pass while ignoring the topographical details. In Fig.~\ref{fig:slope-vs-topo}b (and in the blue-colored line profile in Fig.~\ref{fig:slope-vs-topo}c), a lift height of 120\,nm was chosen when following the slope line mode, resulting in a measurement distance of 60 nm to the top of the structures. The difference between both traces is quite obvious and corresponds well to the simulated traces in Fig.~\ref{fig:slope-vs-topo}d. The non-symmetry observed in the experimental data is attributed to tip tilt, which could be corrected in qMFM. This shows that the clear distortion introduced by following the pronounced sample topography can be fully compensated by following the average topography slope.
        
\section{\label{sec:concl}Summary and Outlook}
       
    While Q-control is a useful feature for performing amplitude-controlled topography measurements in vacuum AFM, its limitations and non-linear behavior in phase shift measurements make it unsuitable for vacuum MFM. However, these challenges can be overcome by measuring frequency shift instead. To address this, the new \textcolor{reviewer2}{high quality-factor two-pass mode} is introduced, combining the advantages of both methods into a fast and sensitive vacuum MFM mode capable of handling magnetic samples with topographical features.
    
    This novel operation mode is implemented via a microcontroller that switches the required signals using CMOS multiplexers to an external lock-in amplifier, allowing frequency shift measurements using a phase-locked loop.
    
    The enhanced sensitivity of this new operation mode is demonstrated by MFM measurements on a nano-patterned magnetic sample. The linearity of the measurement technique is verified using a well-characterized, calculable multilayer reference sample that forms a domain pattern structure. \textcolor{reviewint}{Topography interplay in two-pass mode, as well as a method of circumventing it, is evaluated through measurements on the nano-patterned magnetic sample.}
        
    With this approach, high-sensitivity linear MFM measurements are now feasible on both structured and flat samples, utilizing the standard two-pass MFM technique with frequency-based evaluation. The high-quality two-pass mode paves the way for high-sensitivity, high-resolution quantitative magnetic force microscopy in vacuum, accessible to a wide user base with minimal modifications and retraining. \textcolor{SS}{In the long term, standardization of quantitative vacuum MFM, as recently achieved for ambient qMFM \cite{IEC2021}, would be desirable to further promote industrial applications.}

\section*{\label{sec:acknow}Acknowledgements}
    
    This project was supported by the German Federal Ministry of Economic Affairs and Climate Action (BMWK - Bundesministerium für Wirtschaft und Klimaschutz) within the TransMeT project \textit{"Realisierung eines quantitativen Magnetkraftmikroskopie-Messverfahrens gemäß IEC TS 62607-9-1 mit einem kommerziellen System"}. \textcolor{SS}{We thank Marcelo Jaime and Jantje Kalin for critically reading the manuscript.}

\section*{\label{sec:declarations}Author Declarations}

    \subsection*{\label{sec:confl}Conflict of Interest}
        The authors declare no conflict of interest.

    \subsection*{\label{sec:contrib}Author Contributions}
        \textbf{Christopher Habenschaden:} 
        Conceptualization (equal), 
        Data curation (lead),
        Formal analysis (equal), 
        Investigation (lead),
        Methodology (equal),
        Software (lead),
        Validation (equal),
        Visualization (lead),
        Writing - original draft (lead).
        
        \textbf{Sibylle Sievers:}          
        Conceptualization (lead),
        Data curation (equal),
        Formal analysis (equal),
        Funding acquisition (lead),
        Investigation (equal),
        Methodology (equal),
        Project administration (lead), 
        Resources (lead),
        Software (equal),
        Supervision (lead),
        Validation (equal),
        Visualization (equal),
        Writing - review \& editing (lead).
        
        \textbf{Alexander Klasen:} 
        Data curation (supporting),
        Methodology (supporting),
        Resources (supporting),
        Validation (supporting),
        Writing - review \& editing (supporting).
        
        \textbf{Andrea Cerreta:} 
        Methodology (supporting),
        Resources (supporting),
        Validation (supporting).
        
        \textbf{Hans Werner Schumacher:} 
        Conceptualization (equal),
        Formal analysis (equal),
        Funding acquisition (equal),
        Methodology (equal),
        Project administration (equal), 
        Resources (equal),
        Supervision (equal),
        Validation (equal),
        Writing - review \& editing (equal).

\section*{\label{sec:data}Data availability}

     	The data that support the findings of this study are available from the corresponding author upon reasonable request.

\section*{References}

\bibliography{two-pass-dual-mode_bib.bib}

\end{document}